\documentclass[twocolumn,aps,superscriptaddress]{revtex4-1} 
\usepackage{graphicx}
\usepackage{amsmath, amsthm, amssymb}
\usepackage{epstopdf}
\usepackage{mathtools}
\usepackage{graphicx}
\usepackage{dcolumn}
\usepackage[tight]{subfigure}
\usepackage{amsmath}
\usepackage{verbatim}
\usepackage{inputenc}
\usepackage{color}
\usepackage{bm} 
\usepackage{natbib}
\usepackage{xspace}
\usepackage{mathtools}
\usepackage{hyperref} 
\usepackage{mathbbol}

\renewcommand{\vec}[1]{\mathbf{#1}}

\begin{document}
\preprint{APS/123-QED}

\title{Josephson Quantum Heat Engine}
\author{G. Marchegiani}
\email{giampiero.marchegiani@df.unipi.it} 
\affiliation{Dipartimento di Fisica dell'Universit\`{a} di Pisa, Largo Pontecorvo 3, I-56127 Pisa, Italy}
\affiliation{NEST Istituto Nanoscienze-CNR and Scuola Normale Superiore, I-56127 Pisa, Italy}

\author{P. Virtanen}
\affiliation{NEST Istituto Nanoscienze-CNR and Scuola Normale Superiore, I-56127 Pisa, Italy}

\author{F. Giazotto}
\email{francesco.giazotto@sns.it} 
\affiliation{NEST Istituto Nanoscienze-CNR and Scuola Normale Superiore, I-56127 Pisa, Italy}

\author{M. Campisi}
\affiliation{NEST Istituto Nanoscienze-CNR and Scuola Normale Superiore, I-56127 Pisa, Italy}

\date{\today}

\begin{abstract}
The design of a mesoscopic \textit{self-oscillating} heat engine that works thanks to purely quantum effects is presented.
The proposed scheme is amenable to experimental implementation with current state-of-the-art nanotechnology and materials. One of the main features of the structure is its versatility: The engine can deliver work to a generic load without galvanic contact. This makes it a promising building block for low-temperature on-chip energy management applications.
The heat engine consists of a circuit featuring a thermoelectric element based on a ferromagnetic insulator-superconductor tunnel junction and a Josephson weak link that realizes a purely quantum DC/AC converter. This enables contactless transfer of work to the load (a  generic RL circuit). The performance of the heat engine is investigated as a function of the thermal gradient applied to the thermoelectric junction. Power up to $1$ pW can be delivered to a load $R_L=10\ \Omega$. 
\end{abstract}

\maketitle

One question that is currently in the  limelight of intense theoretical investigation is whether and to what extent quantum effects may play a role in the performance of nano-scale heat engines \cite{uzdin_equivalence_2015}. While it is clear that the laws of thermodynamics hold unaltered in the microscopic realm, even when a quantum description is necessary \cite{alicki_quantum_1979,campisi_fluctuation_2014} it is widely recognized that quantum effects can have an impact  on an engine's performance nonetheless \cite{scully_extracting_2003,scully_quantum_2011,uzdin_equivalence_2015,rahav_heat_2012,mitchison_coherence-assisted_2015}. Whether such impact is positive depends largely on the type of engine design that one considers and, as well, on what quantity one is interested in (e.g. power output vs. efficiency) \cite{campisi_dissipation_2016}.

The recent quest for solid state quantum technologies, with the rise of coherent caloritronics \cite{giazotto_josephson_2012,martinez-perez_coherent_2014,jose_martinez-perez_quantum_2014,giazotto_phase-controlled_2012,fornieri_nanoscale_2016,giazotto_opportunities_2006}, phononics \cite{li_textitcolloquium_2012} and the latest development in spintronics \cite{linder_superconducting_2015} demand for the implementation of a versatile mesoscopic engine working at cryogenic temperatures. Superconducting devices hold a promise in this respect, but heat engines based on superconducting elements are almost unexplored in the literature. One experimental proposal has been put forward in Ref. \cite{niskanen_information_2007} where a driven superconducting qubit couples alternatively to two resistors kept at different temperatures, thus realizing a quantum Otto engine/refrigerator. A similar proposal uses two qubits operated by a  quantum SWAP gate \cite{campisi_nonequilibrium_2015}. These studies, like many theoretical studies in the field, disregard the impact that the load can have on their performance and as well on possible mechanisms of persistent steady work extraction \cite{alicki_thermoelectric_2016-1}. An interesting recent proposal where the the load is explicitly accounted for, uses photon assisted tunneling to make Cooper-pairs in a superconducting device climb up a voltage gradient \cite{hofer_quantum_2016}. In that case the load is not generic, but rather specific.

\begin{figure}[tp]
\begin{centering}
\includegraphics[width=0.5\textwidth]{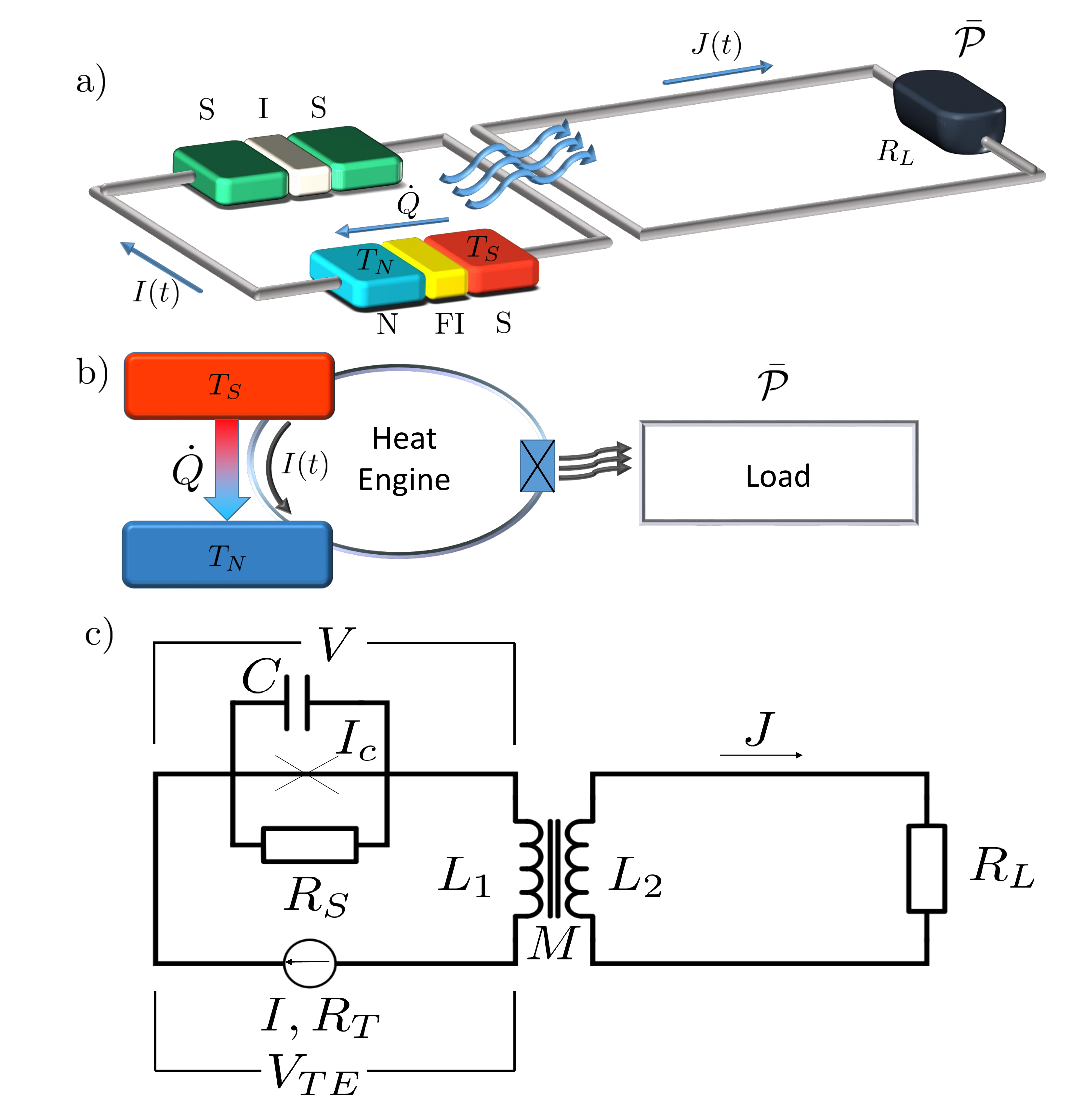}
\caption{(color online). a) Scheme of the Josephson quantum heat engine. b) Thermodynamic scheme. The heat current generates a DC electric current, which is (partly) converted into an AC signal by the Josephson junction. A corresponding AC current is induced to the load circuit by electromagnetic coupling. c) Equivalent electric circuit. The symbols and the voltage convections used in the text are indicated.}
\label{Fig1}
\end{centering}
\end{figure}
Here we present the design and quantitative analysis of a mesoscopic heat engine that i) sustains persistent work extraction thanks to self-oscillations whose origin is purely quantum mechanical (i.e., the Josephson effect) ii) is amenable to experimental implementation with current state-of-the-art nanotechnology, iii) is highly versatile as it can transfer work to a generic load without galvanic contact. The envisaged application of the engine is in the realm of low-temperature on-chip heat management as an energy harvesting block. The design can be seamlessly integrated within a coherent caloritronics platform \cite{giazotto_josephson_2012,martinez-perez_coherent_2014}. Since the engine would not work in the absence of quantum phenomena, it embodies an example where quantum effects provide an evident advantage. Below we show that, for realistic experimental parameters, the structure can deliver a remarkable amount of power to a generic load.

The Josephson quantum heat engine is schematically illustrated in Fig. \ref{Fig1}. The heat engine consists of a closed circuit featuring a \textit{N}-FI-S (Normal metal-Ferromagnetic Insulator-Superconductor) tunnel junction connected via superconducting wires to a Josephson weak link, forming the primary circuit. The load is a generic LR circuit, the secondary circuit. Engine and load are coupled via electromagnetic induction. 

As recently predicted \cite{ozaeta_predicted_2014} and experimentally demonstrated \cite{kolenda_observation_2016} the spin-filtered tunnel junction to spin-split superconductor features a very large thermoelectric effect in the presence of a magnetic field. Such thermoelectric effect originates from a purely quantum-mechanical property, namely electron spin. The N-FI-S tunnel junction has been proposed for electron cooling of the normal metal \cite{kawabata_heat_2015}, as a sensitive thermometer \cite{giazotto_ferromagnetic-insulator-based_2015} and as a key element for a phase-coherent thermoelectric transistor \cite{giazotto_proposal_2014}.  When a temperature difference $\delta T=T_S-T_N$ is applied to the $N$-FI-S element a thermocurrent flows in the primary circuit. If it is strong enough to drive the Josephson junction into the resistive state, an AC current with frequency $\nu=|V|/\Phi_0$ is superimposed in the circuit  (here $V $ is the junction voltage and $\Phi_0\simeq 2.067\times10^{-15}\ \textrm{Wb}$ is the flux quantum). This electromagnetically induces an AC current in the secondary circuit.

\emph{Engine-Load modelling.--}
The engine features a \textit{N}-FI-S junction, working as a thermoelectric element.
The key role is played by the FI layer, which breaks the particle-hole symmetry of the superconductor density of states $N_{\uparrow,\downarrow}$ of the spin up ($\uparrow$) and spin down ($\downarrow$) bands, by inducing an exchange field $h_{exc}$ in the superconductor S
\begin{equation}
N_{\uparrow,\downarrow}(E)=\frac{1}{2}\left|{\mathrm{Re}\left[\frac{E+i\Gamma\pm h_{exc}}{\sqrt{(E+i\Gamma\pm h_{exc})^2-\Delta^2}}\right]}\right|.
\end{equation}
Furthermore the FI layer acts as a spin filter with polarization $P=(G_\uparrow-G_\downarrow)/(G_\uparrow+G_\downarrow)$ \cite{moodera_phenomena_2007}, where $G_{\uparrow,\downarrow}$ is the spin up (down) junction normal state conductance.
Here $\Gamma$ is the phenomenological broadening parameter \cite{dynes_tunneling_1984} of the BCS density of states and $\Delta(h_{exc},T_S)$ is the pairing superconducting potential which is computed in a self consistent way \cite{giazotto_ferromagnetic-insulator-based_2015,giazotto_superconductors_2008}. We assume that the $S$ layer is thinner than the superconducting decay length of the exchange interaction. In this approximation the exchange field $h_{exc}$ is spatially homogeneous. 

In the tunneling limit, the DC current flowing through the $N$-FI-S element is given by \cite{ozaeta_predicted_2014}
\begin{equation}
I_{TE}(V_{TE})\text=\frac{1}{e R_T}\underset{-\infty}{\overset{\infty}{\mathop \int }}dE\mathcal N(E)[f_S(T_S)\text-f_N(V_{TE},T_N)]
\label{eq:thermocurrent}
\end{equation}
where $-e$ is the electron charge, $R_T$ is the normal-state resistance of the junction and $V_{TE}$ is the bias voltage.
Here $\mathcal N=N_+ + PN_-$, with $N_{\pm}=N_\uparrow \pm N_\downarrow$ and we assume thermal equilibrium states on both sides of the junction  $f_N(V_{TE},T_N)=[1+\exp(E+eV_{TE}/k_B T_N)]^{-1}$ and $f_S(T_S)=[1+\exp(E/k_B T_S)]^{-1}$. Here $T_{S(N)}$ are respectively the temperatures of the $S(N)$ layers. The presence of the exchange field reduces the zero magnetic field critical temperature $T_{C0}$ of the $S$ terminal to the smaller value $T_C<T_{C0}=\Delta_0/1.764 k_B$ \cite{giazotto_superconductors_2008}.

\begin{figure}[tp]
\begin{centering}
\includegraphics[width=0.5\textwidth]{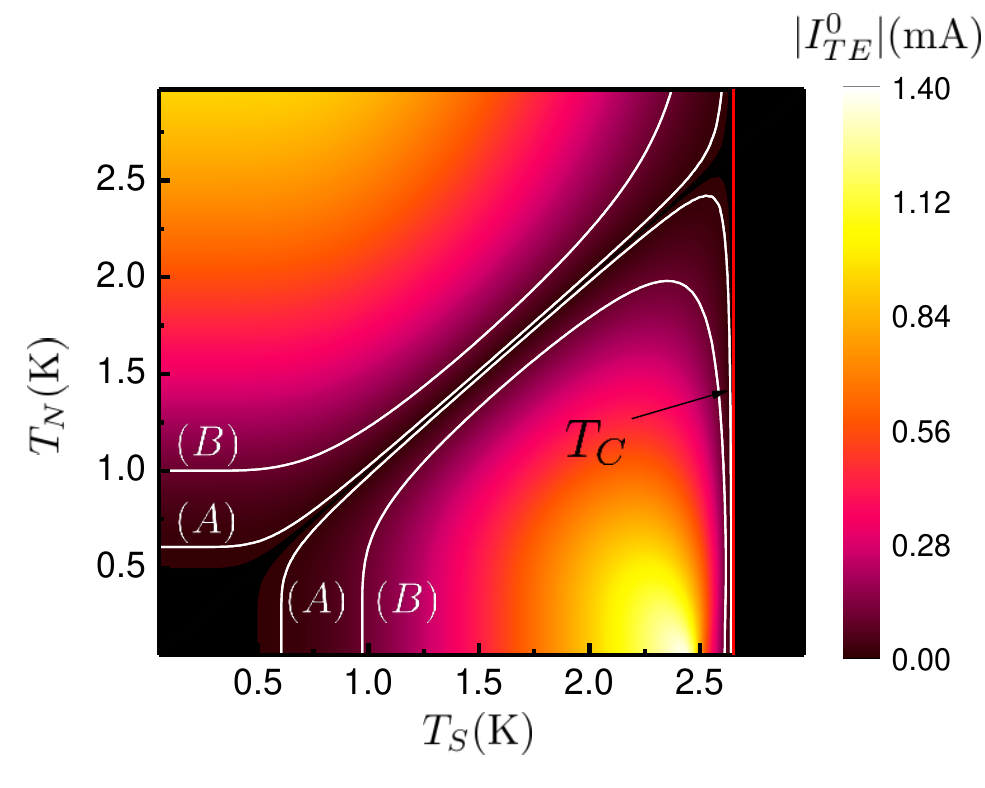}
\caption{(color online). Density plot of zero voltage thermoelectric current provided 
 by the $N$-FI-S element. Parameters are $P=0.9$, $T_{C0}=3.0$ K, $R_T=0.1\ \Omega$, $\Gamma= 10^{-4}\ \Delta_0$ and $h_{exc}=0.4\ \Delta_0$, where $\Delta_0=1.764\ k_B T_{C0}$ and $T_{C0}$ is the zero-field critical temperature. In the black region $|I_{TE}^0|<3\ \mu\textrm{A}$. The contour lines for $|I_{TE}^0|=10\ \mu\textrm{A}$ (A) and $|I_{TE}^0|=100\ \mu\textrm{A}$ (B) are plotted in white.   }
\label{Fig2}
\end{centering}
\end{figure}

Figure \ref{Fig2} shows the absolute value $|I_{TE}^{0}|$ of the thermoelectric current generated by the \textit{N}-FI-S element when its two terminals are kept at the same voltage (i.e. when they are short-circuited via a superconducting connection),
as a function of their temperatures $T_N$ and $T_S$. Two main features can be observed: i) the thermoelectric current is rather small ($< 3\ \mu\textrm A$) when both temperatures are below $T_{min}\simeq 0.5$ K, ii) the thermoelectric current is null when $T_S$ is above the 
critical temperature $T_C\simeq\ 2.6\ \mathrm{K}$ where the $S$-terminal ceases to be superconducting. Accordingly the thermoelectric current is not symmetric \cite{giazotto_ferromagnetic-insulator-based_2015} with respect to the exchange $T_S \leftrightarrow T_N$. \footnote{If $T_N$ is increased at fixed $T_S$, the thermoelectric current increases, as expected. The same is not true in general for increasing $T_S$ at fixed $T_N$. In this case the thermoelectric current increases for small $T_S$, but it then it decreases  on approaching $T_C^*$.}  The strength of the thermoelectric effect increases with $P$ and decreases with $\Gamma$, as shown in \cite{giazotto_ferromagnetic-insulator-based_2015}.

Consider now the case when the two terminals of the \textit{N}-FI-S element are connected via a Josephson junction.
If the Josephson junction stays in the dissipationless regime, the terminals of \textit{N}-FI-S are effectively short-circuited and 
the $DC$ current $I_{TE}^0$ shown in Figure \ref{Fig2} flows in the circuit. In this case no work is done on the load. Hence, a necessary condition for the heat engine to work is that the junction critical current $I_c$ be smaller than $|I_{TE}^{0}|$. As we will show below, in this case a time dependent periodic current flows in the primary circuit, because of the AC Josephson effect, driving a corresponding time dependent current in the secondary circuit, which accordingly receives energy in the form of work.
 
In order to calculate the current $J$ in the load circuit and hence the power dissipated on the load resistor, we 
model the Josephson junction according to the RCSJ model, Eq. (\ref{eq:RCSJ-1},\ref{eq:RCSJ-2}) \cite{stewart_currentvoltage_1968,mccumber_effect_1968}, and write Kirchoff's laws for primary and secondary circuit, Eqs. (\ref{eq:voltagerules-1},\ref{eq:voltagerules-2})
\begin{align}
I &= I_c\sin{\varphi} + \frac{V}{R_{S}} + C\dot{V} \label{eq:RCSJ-1}\\
\dot{\varphi} &=\frac{2\pi}{\Phi_0}V \label{eq:RCSJ-2}\\
{V}_{TE} &= L_1 \dot{I}-M\dot{J}+ V \label{eq:voltagerules-1}\\
M\dot{I} & =L_2\dot{J}+ R_L J\label{eq:voltagerules-2}
\end{align}
where $L_1,L_2$ are the self-inductances of the two circuits, $M$ is their mutual inductance, $R_L$ is the load resistance,  $\varphi$ is the phase difference of the superconducting condensate across the Josephson junction, $V$ is the voltage drop across the Josephson junction and $V_{TE}$ is the $N$ layer voltage, measured with respect to the $S$ layer. The latter depends on the two temperature $T_S$ and $T_N$, and notably, also on the current $I(t)$ in the primary circuit. This occurrence makes the solution of the problem rather involved. The full solution is provided in the Supplementary material. Below we discuss a simplified but quantitatively equivalent solution.

To ease the computation and give a physical picture, we make the Ansatz 
$V_{TE}(t) = \bar V_{TE}+\delta V_{TE}(t)$ with $\bar V_{TE}$ the DC component and $\delta V_{TE}(t)$ a periodic time dependent component. 
We neglect the frequency dependence of the junction impedance $Z(\omega)\simeq Z(0)\simeq R_T$, and write the current in the first circuit as 
\begin{align}
   I(t) \approx -I_{TE}(\bar{V}_{TE}) - \frac{\delta V_{TE}(t)}{R_T}
   \,.
   \label{eq:thermoelectricsimply}
\end{align}
We solve the set (\ref{eq:RCSJ-1},\ref{eq:RCSJ-2},\ref{eq:voltagerules-1},\ref{eq:voltagerules-2},\ref{eq:thermoelectricsimply}) [with $V_{TE}(t) = \bar V_{TE}+\delta V_{TE}(t)$ in (\ref{eq:voltagerules-1})] for $I,J,\varphi,\delta V_{TE},V$, and impose the condition that $\delta V_{TE}$ averages to zero, as from our Ansatz. That is, $\bar V_{TE}$ enters in the equations as a tunable parameter that we determine self consistently. We solve the system of differential equations numerically. The time-dependent component is found to be periodic with natural frequency $\omega=2\pi |\bar V|/\Phi_0$, where $\bar V$ is the average potential drop across the Josephson junction, as expected from the AC Josephson effect.
Notably, for $|I_{TE}(0)|\leq I_c$ the system of equations admits the trivial solution $I=-I^0_{TE},J=0, \varphi=\varphi_0, \delta V_{TE}=0, V=0$ (with some constant $\varphi_0$). That is, if the thermoelectric effect is not strong enough to drive the Josephson junction into the resistive state, the engine delivers null power.

\begin{figure}[tp]
\begin{centering}
\includegraphics[width=0.5\textwidth]{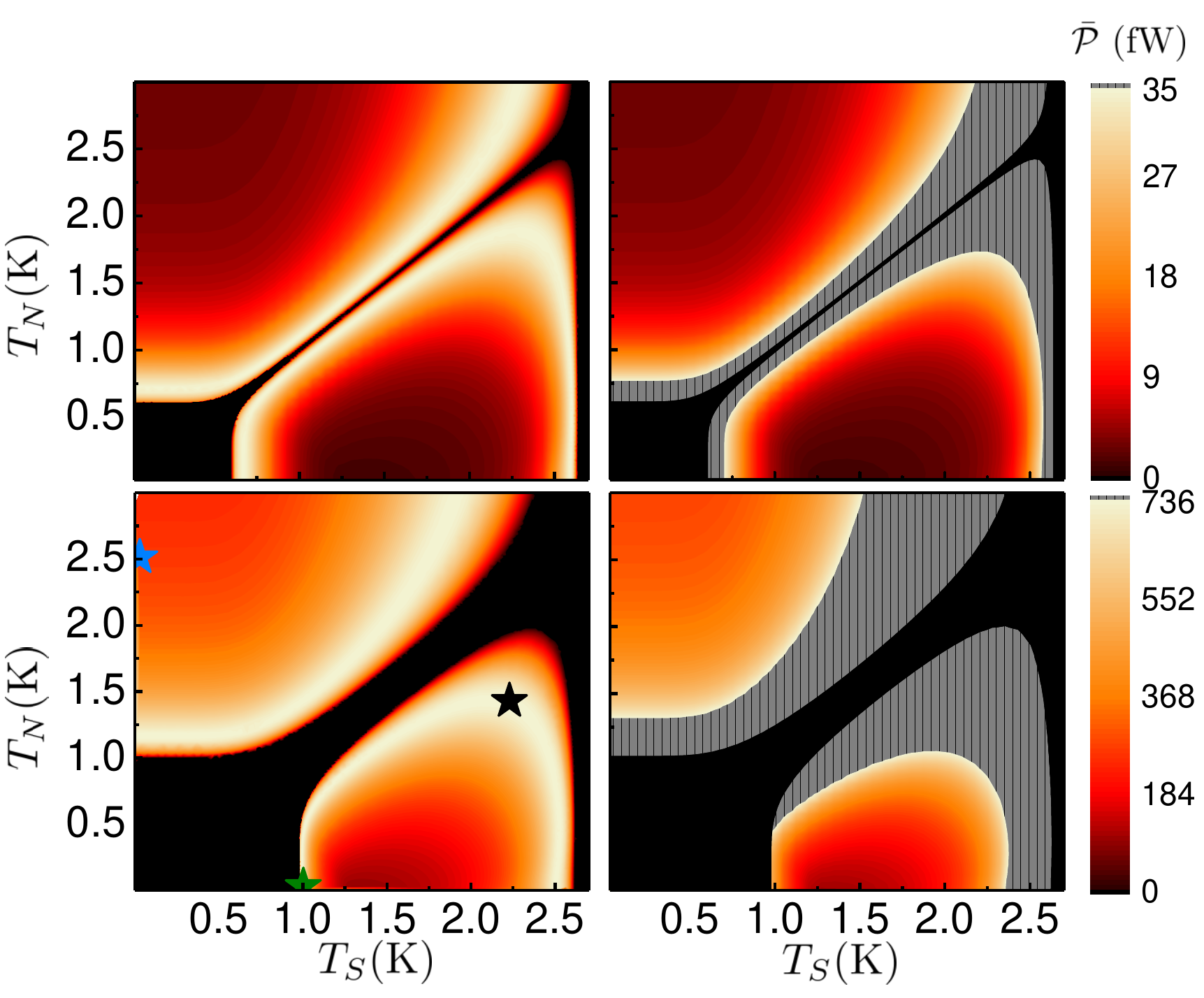}
\caption{(color online). Transmitted power vs thermal gradient for $I_c=10\ \mathrm{\mu A}$ (top) and $I_c=100\ \mathrm{\mu A}$ (bottom). Numerical data (left panels) are compared to the expression Eq. (\ref{eq:powersinusoidal}) (right panels). The regions where Eq. (\ref{eq:powersinusoidal}) gives a higher estimate than the largest value of $\bar{\mathcal{P}}$ in the corresponding numerical graph are drawn in grey. The stars denote the points investigated in Fig. \ref{fig:Fig4}. Parameters are $C=100\ \mathrm{fF}$, $L_1=L_2=100\ \mathrm{pH}$, $M=10\ \mathrm{pH}$, $R_S=1\ \Omega$  and $R_L=10\ \Omega$. }
\label{Fig3}
\end{centering}
\end{figure}

\emph{Power: numerical calculation--} Figure \ref{Fig3} (left panels) shows a contour plot of the average power $\bar{\mathcal{P}}={T^{-1}}\int_{0}^{T} R_L J(t')^2 dt'$ transmitted to the load as a function of the the temperatures $T_S$ and $T_N$ of the normal metal and superconducting layers of the $N$-FI-S element for $I_c =10\ \mathrm {\mu A}$ and $I_c=100\ \mathrm{\mu A}$, as obtained from the self-consistent numerical integration of the set of non-linear equations (\ref{eq:RCSJ-1},\ref{eq:RCSJ-2},\ref{eq:voltagerules-1},\ref{eq:voltagerules-2},\ref{eq:thermoelectricsimply}). 
The plots display a number of interesting features: i) No signal is transmitted in the region $T_S>T_C$ where there is no thermoelectric effect, and in the region $|I_{TE}^0|<I_C$, compare with Fig. \ref{Fig2}, where there is no AC Josephson effect. ii) As $T_N$ is increased, at fixed $T_S$, the power displays a maximum. As $T_S$ is increased, at fixed $T_N$, the power displays  two maxima. Quite interestingly the maxima tend to be close to the boundary between the on and off regions. iii) A stronger thermoelectric effect is not necessarily accompanied by a larger power. iv) Much higher power can be achieved in the case $I_c= 100\ \mathrm{\mu A}$ ($\bar{\mathcal{P}}_{max} \sim 0.74\ \mathrm{pW}$ as compared to the case $I_c= 10\ \mathrm{\mu A}$ ($\bar{\mathcal{P}}_{max} \sim 0.035\ \mathrm{pW}$). 

\emph{Power: analytical estimate.--}
One way to get a rough understanding of the phenomenology displayed in Fig. \ref{Fig3} is to consider the thermoelectric element as an ideal current generator providing some $DC$ current $I_{DC}$, with $|I_{DC}|>I_c$. When the ideal current generator is connected to Josephson junction, a time dependent current is superimposed to the current provided by the generator. In the overdamped limit $C\to 0$, the voltage drop $V(t)$ across the junction (in the absence of any inductive element) can be calculated analytically
\cite{antonio_barone_&_gianfranco_paterno_physics_????}. The solution turns from a sequence of pulses when $|I_{DC}|$ is close to $I_c$ to a sinusoidal signal superimposed to the DC bias, when $|I_{DC}|\gg I_c$, reading $V\simeq I_{DC} R_S + I_c R_S \sin{\omega t}$  with $\omega=2\pi |I_{DC}| R_S/\Phi_0= 2\pi |V_{DC}| /\Phi_0$. In this limit the $AC$ components of the voltage and current are very small compared to the respective $DC$ components. Therefore the DC voltage can be calculated as solution of  $I_{TE}(V_{DC})=-V_{DC}/R_S$. Accordingly, one can see the thermoelectric element-Josephson junction  series as an AC voltage generator with internal resistance $R_T$, and the Kirchoff's relations become
\begin{equation}
I_c R_S \sin{\omega t}\text= R_T\delta I + L_1 \dot{\delta I}-M\dot{J} \label{eq:newprimary}\ ,\ M\dot{\delta I} \text=L_2\dot{J}+ R_L J.
\end{equation}
where $\delta I=-\delta V_{TE}/R_T$ is the AC current in the primary circuit. 
Solving for $J$ the average power in the high frequency limit is given by the simple expression 
\begin{equation}
\bar{\mathcal{P}} =\left(\frac{M I_c R_S}{2L_1}\right)^2\frac{2R_L}{(\omega L_e)^2+R_L^2} 
\label{eq:powersinusoidal}
\end{equation}
with $L_{e}=(L_1L_2-M^2)/L_1$.

Figure \ref{Fig3} (right panels) shows a density plot of the approximate expression (\ref{eq:powersinusoidal}). The grey region denotes the region where Eq. (\ref{eq:powersinusoidal}) gives a higher estimate than the largest value of $\bar{\mathcal{P}}$ in the corresponding numerical graph (left panels). The expression (\ref{eq:powersinusoidal}) provides a better estimation of the numerically computed $\bar{\mathcal{P}}$  when the thermoelectric effect is stronger, as expected on the basis of the discussion above.

Note that on the one hand a strong thermoelectric current is necessary to drive the junction into the resistive state and hence makes the energy transfer to the load possible. On the other hand however a stronger thermoelectric effect means higher driving frequency $\omega$  according to the formula $\omega=2\pi |I_{DC}| R_S/\Phi_0$. Since the impedance of the $RL$ circuit increases with the frequency $\omega$ at some point an increasing thermoelectric effect results in a decreasing power output. This interplay of the two effects is captured by the numerical plots (left panels of Fig. \ref{Fig3}), 
which in fact display the maxima, resulting from their joint action.
The simplified analytical expression instead only captures the latter effect, as it disregards the nonlinear back action of the load on the thermoelectric element, and only captures the linear, inductive features. Note that the analytical formula also captures the fact that the output power increases with the strength of magnetic coupling $M/L_1$ and with the size of the DC-to-AC conversion $I_c R_S$.

\emph{Role of Load.--}
In Fig.4 we investigate the impact of the load on the heat engine performance, for the $(T_S,T_N)$ points marked in Fig.3. We consider i) a point located far from the off region (light blue curve), ii) a point with high output power and medium distance from the off region (black curve), iii) a point with high output power very close to the off region (green curve). All curves show a characteristic resonance peak, which is located at $R_2/\omega L_e$ for the light blue curve, according to Eq. (\ref{eq:powersinusoidal}), and drifts to smaller values by decreasing the distance from the off region. This effect is consistent with the model, since close to the off region, $|\bar{V}_{TE}|$ is significantly smaller than $|V_{DC}|$.

\begin{figure}
\begin{centering}
\includegraphics[width=0.5\textwidth]{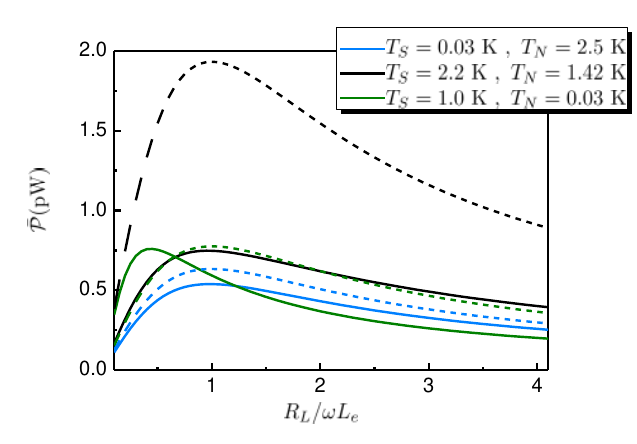}
\caption{(color online). Transmitted power vs load resistance for different working point in the $(T_S,T_N)$ space, marked by the stars in Fig. \ref{Fig3}. Numerical data (solid lines) are compared with the expression Eq.(\ref{eq:powersinusoidal}) (dash lines). Parameters are $C=100\ \textrm{fF}$, $L_1=L_2=100\ \textrm{pH}$, $M=10\ \textrm{pH}$, $R_S=1\ \Omega$,  and $I_c=100\ \mu\textrm{A}$.}
\label{fig:Fig4}
\end{centering}
\end{figure}
\emph{Efficiency.--}
The engine performance in terms of power, which is in the range of the pico-Watt, is rather high as compared to other engines based on superconducting devices operating in the femto-Watt range, e.g. in Ref. \cite{hofer_quantum_2016}. Its performance is however quite low in terms of thermodynamic efficiency. While the device presented in \cite{hofer_quantum_2016} achieves $\eta\sim 0.78$ at maximum power, ours achieves an efficiency that is less than a part per million, $\eta=\bar{\mathcal{P}}/\dot{Q}\lesssim 10^{-6}$, for the parameters used to plot Fig. \ref{fig:Fig4}. Here $\dot{Q}\text=\int_{-\infty}^{+\infty}dE(E+eV_{TE})\mathcal N(E)[f_S(T_S)\text-f_N(V_{TE},T_N)]/(e^2 R_T)$ is the heat current through the thermoelectric element. The reason for such a low efficiency is that, despite the thermoelectric conversion is extremely efficient, most of the the thermoelectrically generated current is dissipated in the shunt resistor $R_S$, and only a little fraction of it is converted into an AC current, via the AC Josephson effect, which drives the secondary circuit, hence delivers power to the load resistor. The high pay for such a low efficiency is compensated not only by the rather high power output, but notably also by the versatility of the design, where the load is very generic, and is addressed wirelessly.

\emph{Practical considerations.--}
All the numerical calculations were performed for parameters realistic for the
implementation of the heat engine. Suitable materials for the N-FI-S junction are europium (Eu) chalcogenides (like EuO, EuS or EuSe) for the FI, providing $P$ up to $\sim98\%$ \cite{moodera_variation_1993,santos_determining_2008}, combined with ultrathin films of superconducting aluminum (Al), with critical temperature $\sim 3\ \mathrm K$ \cite{liu_electrostatic_2013}. Copper or silver are reasonable choices for the N region. The ratio $h_{exc}/\Delta_0$ ranges typically from $0.2$ up to $0.6$ \cite{li_observation_2013}. An alternative to the chalcogenides is to use GdN with superconducting NbN, characterized by the higher critical temperature of NbN of $\sim 15$ K \cite{senapati_spin-filter_2011}. The value for the inductive coupling $M=10\ \mathrm{pH}$ can be achieved with a proper design of the interface between the two circuits. 

\emph{Conclusions.--}
We have shown that a circuit composed by a \textit{N}-FI-S element and a Josephson junction works as a heat engine whose operation rests on genuinely quantum mechanisms.
Based on standard modelling of the \textit{N}-FI-S element and the Josephson junction we have established that the engine is able to transfer without any galvanic contact powers up to $\sim1\ \mathrm{pW}$, which are rather large for mesoscopic superconducting engines. The possibility of switching the operating region of the heat engine by adjusting the Josephson junction critical current can be used to build a power valve
(the critical current can be tuned, for instance, by means of a SQUID \cite{tinkham2004introduction}). Its flexibility combined with the simplicity of the design makes it a promising building block as a power generator at the meso and nanoscale with potential applications in current superconductor-based experimental setups, including coherent caloritronics.

\begin{acknowledgments}
We thank Fabio Taddei, Federico Paolucci and Claudio Guarcello for useful discussions. We acknowledge the European Research Council under the European Union's Seventh Framework Program (FP7/2007-2013)/ERC Grant agreement No. 615187-COMANCHE for partial financial support. The work of M.C. is supported by a Marie Curie Intra European Fellowship within the 7th European Community Framework Programme through the project NeQuFlux grant n.623085 and by the COST action MP1209 â``Thermodynamics in the quantum regime''. The work of P.V. is supported by MIUR-FIRB2013-Project Coca (Grant No. RBFR1379UX).
\end{acknowledgments}

%


\begin{widetext}

\setcounter{equation}{0}
\setcounter{figure}{0}
\setcounter{table}{0}
\setcounter{page}{6}
\makeatletter
\renewcommand{\theequation}{S\arabic{equation}}
\renewcommand{\thefigure}{S\arabic{figure}}
\renewcommand{\bibnumfmt}[1]{[S#1]}
\renewcommand{\citenumfont}[1]{S#1}

\section*{Supplementary Material: Josephson Quantum Heat Engine}
The time-dependent response of a $N$-FI-S junction can be analyzed
via a standard tunneling Hamiltonian calculation.
\cite{werthamer1966-nso,larkin1967-teb,barone1982} Here $e=\hbar=k_B=1$. Retaining spin
labels and considering the S/N case, Eq.~(1) in
Ref.~\onlinecite{werthamer1966-nso} reads,
\begin{equation}
  \label{eq:current}
  I(t)
  =
  -2\Re\sum_{\sigma}\sum_{\vec{k}\vec{q}}\int_{-\infty}^t d{t'}
  e^{\eta t'}|T_{\vec{k}\vec{q}}^\sigma|^2
  e^{i[\phi(t)-\phi(t')]}
  [
    G^<_{\vec{k}\sigma}(t',t)G^>_{\vec{q}\sigma}(t,t')
    -
    G^>_{\vec{k}\sigma}(t',t)G^<_{\vec{q}\sigma}(t,t')
  ]
  \,,
\end{equation}
where $\eta\to0^+$, $T$ is the tunneling matrix element, $G$ are
Green functions for the superconductor ($\vec{k}$) and the normal
($\vec{q}$) sides, and the phase $\phi(t)=\int^t d{t'}V(t')$ is
related to the voltage difference across the junction.  We also assume
for simplicity that tunneling is spin-conserving and that spin
polarizations are collinear. After standard approximations --- assuming
internal equilibrium in the terminals and neglecting momentum dependence
of tunneling matrix element --- we arrive at
\begin{equation}
  I(t)
  =
  \Re\int_{-\infty}^t dt'
  e^{i[\phi(t)-\phi(t')]}
  e^{\eta t'}
  \int_{-\infty}^\infty\frac{dV}{\pi}
  e^{-i(t-t')V} I_{TE}(V)
  \,,
\end{equation}
where $I_{TE}(V)$ is the dc current-voltage relation in Eq.~(2) of the
main text. For $\phi(t)=Vt$, we recover $I(t)=I_{TE}(V)$.

It is useful to add and subtract the Ohmic response $V/R_T$ and the
zero-bias thermoelectric current $I_{TE}(0)$ to remove the singular
parts of the memory kernel. This yields,
\begin{equation}
  I(t)
  =
  \frac{\dot\phi(t)}{R_T}
  +
  I_{TE}(0)
  +
  \int_{-\infty}^t dt'[e^{i[\phi(t)-\phi(t')]} - 1]{\cal K}(t-t')
  \,.
\end{equation}
where
\begin{equation}
  {\cal K}(t)
  =
  \sum_{\sigma=\uparrow/\downarrow}
  \frac{1\pm{}P}{R_T}
  \int_{-\infty}^\infty dE
  [N_\sigma(E) - \frac{1}{2}]
  \int_{-\infty}^\infty\frac{dV}{\pi} e^{-iV t}
  [\frac{1}{2} - f_N(E,V,T_N)]
  \,.
\end{equation}
Evaluating the integrals we obtain for $t\ge0$,
\begin{equation}
  {\cal K}(t)
  =
  \frac{1}{R_T}
  \frac{\pi{}T_N\Delta J_1(t\Delta)}{\sinh(\pi T_N t)}
  [i\cos(h t) + P \sin(h t)]
  \,,
\end{equation}
where $J_1$ is a Bessel function.  This memory kernel is well-behaved:
it is bounded, decays as $\sim{}e^{-\pi{}T_Nt}/t^{3/2}$ at large
times, and has no oscillations on time scales shorter than $1/\Delta$,
$1/h$.  We can also note the sum rule,
\begin{equation}
  I_{TE}(0) = \Re{}\int_0^{\infty}dt[{\cal K}(t) - {\cal K}(t)\rvert_{T_N\mapsto{}T_S}]
  \,,
\end{equation}
that connects the thermoelectric dc and ac responses.
Finally, for $V(t) = \bar{V} + \delta V(t)$,
\begin{equation}
  \label{eq:suppl-exact}
  I(t)=
  I_{TE}(\bar{V})
  +
  \frac{\delta{}V(t)}{R_T}
  + \Re\int^t_{-\infty}dt'
  [e^{i\phi(t)-i\phi(t')}-e^{i\bar{V}(t-t')}]{\cal K}(t-t')
  \,.
\end{equation}
The value of $\bar{V}$ can be chosen arbitrarily.  In the simplified
model in the main text, we choose $\bar{V}=\overline{\dot{\phi}}$
self-consistently and discard the time-nonlocal term.

Replacing Eq. (S-7) with Eq. (7) in the main text leads to a set of non linear differential equation that we solve numerically, without the need to enforce self-consistency: the new set is automatically self-consistent. The new solution excellently agrees with the solution shown in Fig. 3 left panels, apart for a small region at low $T_N$ around the lower $T_S$ peak, where the full fledged solution predicts higher power. 

\end{widetext}
\end{document}